\documentclass{article}
\usepackage{fullpage}

\usepackage{amsmath}
\usepackage{amsthm}
\usepackage{amssymb}
\usepackage{amstext}
\usepackage{graphicx}
\usepackage{stmaryrd}

\usepackage{hyperref} 

\usepackage{color}
\usepackage[applemac]{inputenc}

\usepackage{pict2e}

\usepackage{xparse}

\makeatletter
\DeclareRobustCommand{\lintprod}{%
  \mathbin{\mathpalette\int@prod{(0,0)(0.8,0)(0.8,0.6)}}%
}
\DeclareRobustCommand{\rintprod}{%
  \mathbin{\mathpalette\int@prod{(0.1,0.6)(0.1,0)(0.9,0)}}}

\newcommand{\int@prod}[2]{%
  \begingroup
  \sbox\z@{$\m@th#1+$}%
  \setlength\unitlength{\wd\z@}%
  \linethickness{0.09\unitlength}%
  \begin{picture}(1,1)
  \roundcap
  \polyline#2
  \end{picture}%
  \endgroup
}
\makeatother

\usepackage[retainorgcmds]{IEEEtrantools}

\usepackage{tikz}

\newcommand{\Bb}{{\mathbf B}}

\newcommand{\Eb}{{\mathbf E}}

\newcommand{\ebf}{{\mathbf e}}

\newcommand{\jb}{{\mathbf j}}

\DeclareDocumentCommand \spinori { o } {%
  \IfNoValueTF {#1} {%
    \sigma%
  }{%
    \sigma_{#1}%
  }%
}

\DeclareDocumentCommand \indG { o o } {%
  \IfNoValueTF {#1} {%
    \iota%
  }{%
    \IfNoValueTF {#2} {%
      \iota_{#1}%
    }{%
      \iota_{#1}(#2)%
    }
  }%
}

\DeclareDocumentCommand \indGc { o o } {%
  \IfNoValueTF {#1} {%
    \bar\iota%
  }{%
    \IfNoValueTF {#2} {%
      \bar\iota_{#1}%
    }{%
      \bar\iota_{#1}(#2)%
    }
  }%
}

\DeclareDocumentCommand \indGinv { o o } {%
  \IfNoValueTF {#1} {%
    \bar\iota%
  }{%
    \IfNoValueTF {#2} {%
      \iota^{-1}_{#1}%
    }{%
      \iota^{-1}_{#1}(#2)%
    }
  }%
}

\DeclareDocumentCommand \coefG { o o } {%
  \IfNoValueTF {#1} {%
    \gamma%
  }{%
    \IfNoValueTF {#2} {%
      \gamma_{#1}%
    }{%
      \gamma_{#1}(#2)%
    }
  }%
}

\newcommand{\abf}{\mathbf{a}}

\DeclareDocumentCommand \act { o } {%
  \IfNoValueTF {#1} {%
    \mathcal S%
  }{%
    \mathcal S_{\text{#1}}%
  }%
}

\DeclareDocumentCommand \ld { o } {%
  \IfNoValueTF {#1} {%
    \mathcal L%
  }{%
    \mathcal L_{\text{#1}}%
  }%
}

\newcommand{\deltabf}{{\boldsymbol{\partial}}}

\newcommand{\xbpert}{\pmb{\varepsilon}}

\newcommand{\Gb}{\mathbf{G}}
\makeatletter
\def\trMs{\@ifnextchar[{\@with}{\@without}}
\def\@with[#1]{\Gb_{\xbpert}^{#1}}
\def\@without{\Gb_{\xbpert}^s}
\makeatother

\newcommand{\len}[1]{\lvert#1\rvert}

\newcommand{\hodge}{{\scriptscriptstyle\mathcal{H}}}
\newcommand{\hodgeinv}{{\scriptscriptstyle\mathcal{H}^{-1}}}

\newtheoremstyle{question}
  {\topsep}   
  {\topsep}   
  {\upshape}  
  {0pt}       
  {\itshape}  
  {.}         
  {5pt plus 1pt minus 1pt} 
  {\thmname{#1} \thesection.\thmnumber{\itshape#2}\thmnote{(#3)}} 

\makeatletter
    \def\@endtheorem{\hfill$\P$\endtrivlist\@endpefalse }
\makeatother
\theoremstyle{question}

\makeatletter
\let\save@mathaccent\mathaccent
\newcommand*\if@single[3]{%
  \setbox0\hbox{${\mathaccent"0362{#1}}^H$}%
  \setbox2\hbox{${\mathaccent"0362{\kern0pt#1}}^H$}%
  \ifdim\ht0=\ht2 #3\else #2\fi
  }
\newcommand*\rel@kern[1]{\kern#1\dimexpr\macc@kerna}
\newcommand*\widebar[1]{\@ifnextchar^{{\wide@bar{#1}{0}}}{\wide@bar{#1}{1}}}
\newcommand*\wide@bar[2]{\if@single{#1}{\wide@bar@{#1}{#2}{1}}{\wide@bar@{#1}{#2}{2}}}
\newcommand*\wide@bar@[3]{%
  \begingroup
  \def\mathaccent##1##2{%
    \let\mathaccent\save@mathaccent
    \if#32 \let\macc@nucleus\first@char \fi
    \setbox\z@\hbox{$\macc@style{\macc@nucleus}_{}$}%
    \setbox\tw@\hbox{$\macc@style{\macc@nucleus}{}_{}$}%
    \dimen@\wd\tw@
    \advance\dimen@-\wd\z@
    \divide\dimen@ 3
    \@tempdima\wd\tw@
    \advance\@tempdima-\scriptspace
    \divide\@tempdima 10
    \advance\dimen@-\@tempdima
    \ifdim\dimen@>\z@ \dimen@0pt\fi
    \rel@kern{0.6}\kern-\dimen@
    \if#31
      \overline{\rel@kern{-0.6}\kern\dimen@\macc@nucleus\rel@kern{0.4}\kern\dimen@}%
      \advance\dimen@0.4\dimexpr\macc@kerna
      \let\final@kern#2%
      \ifdim\dimen@<\z@ \let\final@kern1\fi
      \if\final@kern1 \kern-\dimen@\fi
    \else
      \overline{\rel@kern{-0.6}\kern\dimen@#1}%
    \fi
  }%
  \macc@depth\@ne
  \let\math@bgroup\@empty \let\math@egroup\macc@set@skewchar
  \mathsurround\z@ \frozen@everymath{\mathgroup\macc@group\relax}%
  \macc@set@skewchar\relax
  \let\mathaccentV\macc@nested@a
  \if#31
    \macc@nested@a\relax111{#1}%
  \else
    \def\gobble@till@marker##1\endmarker{}%
    \futurelet\first@char\gobble@till@marker#1\endmarker
    \ifcat\noexpand\first@char A\else
      \def\first@char{}%
    \fi
    \macc@nested@a\relax111{\first@char}%
  \fi
  \endgroup
}
\makeatother

\newcounter{aside}
   {\refstepcounter{aside}
   \begin{adjustwidth}{\parindent}{-1.5\parindent}
	\small \textbf{Aside \thechapter.\theaside~#1}
	}
{\par\end{adjustwidth}}

\usepackage{xcolor}
\usepackage{color}
\usepackage{amsmath}
\usepackage{amssymb}
\usepackage{geometry}

\usepackage{authblk}
\newcommand\blfootnote[1]{%
  \begingroup
  \renewcommand\thefootnote{}\footnote{#1}%
  \addtocounter{footnote}{-1}%
  \endgroup
}

\newcommand{\jbf}{{\mathbf j}}
\newcommand{\xbf}{{\mathbf x}}

\newcommand{\Iset}{{\mathcal{I}}}

\newcommand{\Jq}{{\textsc{J}}}
\newcommand{\Fq}{{\textsc{F}}}
\newcommand{\Aq}{{\textsc{A}}}
\newcommand{\Dq}{{\textsc{D}}}
\newcommand{\qtf}{{\textsc{Q}}}
\newcommand{\iq}{{\text{i}}}
\newcommand{\jq}{{\text{j}}}
\newcommand{\kq}{{\text{k}}}
\newcommand{\ptf}{{\textsc{P}}}

\newcommand{\Qset}{{\mathcal{Q}}}
\newcommand{\wedgevec}{{\wedge_{\text{vec}} }}
\newcommand{\wedgequat}{{\wedge_{\Qset} }}

\newcommand{\pbf}{\mathbf{p}}
\newcommand{\qbf}{\mathbf{q}}

 \theoremstyle{definition}

 \theoremstyle{definition}

\title{On exterior-algebraic quaternions \\ with application to Maxwell equations
\blfootnote{Published on: {\bf Acta Appl Math}, 198:1 (2025). DOI: \href{https://doi.org/10.1007/s10440-025-00734-w}{10.1007/s10440-025-00734-w}.}}
\author{Ivano Colombaro}
\date{}

\affil{\centering\small Faculty of Engineering, Free University of Bozen-Bolzano, via Bruno Buozzi 1, 39100, Bolzano, Italy 
\newline
\textbf{e-mail:} ivano.colombaro@unibz.it}

\begin{document}

\maketitle

\begin{abstract}
Within the framework of exterior algebra, the concept of time-like quaternions has been previously established. This paper advances beyond the existing structure by elucidating the procedure for constructing time-like quaternions with the feature of complex scalar terms. We delineate on the crucial role of these extended definition of quaternions in formulating Maxwell equations, having properly defined a pure quaternion containing the components of the classical electric and magnetic fields. Through a formal introduction, we describe the approach followed to acquiring time-like quaternions, characterized by having complex scalar term, and their significant relationship with the derivation of Maxwell equations.
This topic not only underscores the mathematical intricacies of quaternionic algebra, but also highlights its profound implications in the description of fundamental electromagnetic phenomena.

\vspace{5mm}
\end{abstract}


\section{Introduction}

Quaternions are interesting mathematical objects that have found applications in various fields of science including computer graphics, robotics, aerospace engineering, and quantum mechanics. Despite their complexity, they offer a powerful way to represent rotations and orientations in three-dimensional space and, for this reason, they find fertile ground in the field geometric algebra~\cite{PerezGracia2020} and in the description of generalized hypercomplex manifolds~\cite{Wells1980,salamon1986diffgeomquatman}.

At their core, quaternions are an extension of complex numbers and they are typically written in the form~\cite{hamilton1853LectureNotes}
\begin{equation}\label{eq:introeq}
    \qtf = a + b\ebf_\iq + c\ebf_\jq + d\ebf_\kq ,
\end{equation}
where $a$ is the scalar part, while $b$, $c$ and $d$ are the components of the vectorial part and the three vector basis elements form an orthogonal basis such that $\ebf_\iq^2 = \ebf_\jq^2 = \ebf_\kq^2 = \ebf_\iq \ebf_\jq \ebf_\kq = -1$~\cite{girard1984quatgroup}.
In a first stage, they were indeed a way to extend complex numbers to higher dimensions~\cite{hamilton1844onquaternions} and they have been later incorporated inside what is called nowadays as Clifford algebra~\cite[Sec.~1.3]{Ablamowicz2004Clifford}.
From a more formal perspective, quaternions are characterized for being a division ring but not a field.
Indeed, since they satisfy the properties of a ring but do not satisfy the property of commutativity for multiplication, they are an example of a non-commutative
division ring~\cite{Fraleigh2003algebra}.

These mathematical structures have been widely studied and formalised under several points of view in several mathematical backgrounds.
One remarkable example might be given by dealing with quaternions in the context of differential geometry, e.~g.~\cite{salamon1986diffgeomquatman, giardino2024diffgeom}.
Another recurring application in the literature deals with the representation of rotation matrices and operators, as could be deduced by~\cite{Kuipers1999quaternions} or~\cite[Sec.~11-2]{Hearn1997ComputerGraphics}, and several other references.


In~\cite{colombaro2023quaternions}, quaternions have been defined by means of the exterior-algebraic formalism, in particular considering the significant example about the efficient representation of three-dimensional rotations.
In this paper, the time-like definition of quaternions formalised in the framework of exterior algebra is considered, with complex scalar part, and considering the spacetime exterior calculus in~\cite{colombaro2019introductionSpaceTimeExteriorCalculus}. 
As a relevant application, spacetime exterior calculus might be rather convenient as a notation for expressing the classical theory of electromagnetism, in a similar way to~\cite{dressel2015stalgebra} dealing with spacetime algebra and to~\cite{girard1984quatgroup} in the context of the quaternion group. 

In particular, we take into account the generalized formulation of Maxwell equations in exterior algebra, written~\cite{colombaro2020generalized}
\begin{gather}
    \deltabf \lintprod \mathbf{F} = \mu_0 \mathbf{J} \label{eq:eamax-inh},\\
    \deltabf \wedge \mathbf{F} = 0 , \label{eq:eamax-h}
\end{gather}
where the operations of left interior product $\lintprod$ and exterior product $\wedge$ are formally defined in the next section. Generalized Maxwell equations~\eqref{eq:eamax-inh} and~\eqref{eq:eamax-h} work for arbitrary spacetime dimensions, indicating conventionally $k$ as time dimensions and $n$ as spatial dimensions~\cite{colombaro2019introductionSpaceTimeExteriorCalculus}.
There exists then an equivalence between quaternions and the generalized formulation of electromagnetism, fixing the spacetime dimensions to $k=1$ and $n=3$ and for the multivector $\displaystyle \mathbf{F} = \frac{E_1}{c} \ebf_{01} + \frac{E_2}{c} \ebf_{02} + \frac{E_3}{c} \ebf_{03} + B_1 \ebf_{23} - B_2 \ebf_{13} + B_3 \ebf_{12}$, given $E_i$ as the components of the electric and $B_i$ as the components of the magnetic flux density, sometimes simply called magnetic field, for $i=1,2,3$, and for the vector $\mathbf{J} = c\rho \ebf_0 + \jb $~\cite[Sec.~3.1]{colombaro2020generalized}, where $c$ is the speed of light, $\rho$ is the density of the total electric charge and $ \jb $ contains the density if the electric current. As a matter of fact, these latter are analogous to the standard Maxwell equations expressed in terms of electric and magnetic field, $\Eb$ and $\Bb$ respectively, namely~\cite{jackson1999classicalElectrodynamics}
\begin{gather}
    \nabla\cdot\Eb = \frac{\rho}{\varepsilon_0} \label{eq:maxEq_intro_1} \\
        \nabla\cdot\Bb = 0 \label{eq:maxEq_intro_2} \\
    \nabla\times\Eb = -\frac{\partial\Bb}{\partial t} \label{eq:maxEq_intro_3} \\
    \nabla\times\Bb = \mu_0\jb + \mu_0\varepsilon_0\frac{\partial\Eb}{\partial t}. \label{eq:maxEq_intro_4}
\end{gather}
since~\eqref{eq:eamax-inh} leads to~\eqref{eq:maxEq_intro_1} and~\eqref{eq:maxEq_intro_4}, while~\eqref{eq:eamax-h} recall to~\eqref{eq:maxEq_intro_2} and~\eqref{eq:maxEq_intro_3}. Conventionally, we employ the \textit{International System of Units}, so that the constants are set accordingly~\cite{PDG2020}.

Hereinafter, in Sec.~\ref{sec:EA}, a concise introduction presenting the main elements and properties of spacetime exterior algebra and calculus are presented, in order to set the framework up, and the formulation of quaternions by means of this notation is presented.
Subsequently, we describe how such structure might be used in order to define Maxwell equations in Sec.~\ref{sec:ME}. 
Finally, we conclude the paper by discussing the obtained results in Sec.~\ref{sec:last} and comparing them with the existing literature.

\section{Mathematical background of exterior algebra and calculus}
\label{sec:EA}

Spacetime exterior calculus adopted in this work was firstly introduced in~\cite{colombaro2019introductionSpaceTimeExteriorCalculus}, where the formalism and the main properties have been set up together with the algebraic characteristics and the concept of multivectors. 
Dealing with exterior algebra involves the use of the so called exterior product, also known as wedge product, represented by the symbol $\wedge$.
In a $(k,n)$ spacetime, we define the canonical basis $\displaystyle{\{\ebf_i\}_{i=0}^{k+n-1}}$, having conventionally the first $k$ elements labelled from $0$ to $k-1$ as the time-like coordinates and to the following indices, until $n-1$, as the space-like coordinates. 

The $\wedge$ product applied to such basis elements delineate the $s$-graded basis that characterizes multivectors, namely
\begin{equation}\label{eq:mvpI}
    \ebf_I = \ebf_{i_1}\wedge \ebf_{i_2}\wedge \dots \wedge \ebf_{i_s} ,
\end{equation}
where the list $I=(i_1, i_2 , \dots i_s)$ refers to non-repeated ordered indices.
As a consequence, calling as $\Iset_s$ the set of all the lists $I$ satisfying~\eqref{eq:mvpI}, we write a multivector field as
\begin{equation}
    \mathbf{V} = \sum_{I\in \Iset_s} V_I \ebf_I .
\end{equation}
The exterior product, as introduced previously, can also be defined between multivectorial basis elements of different degree. Indeed, taking an $s$-vector basis $\ebf_I$ with degree $s=\len{I}$ and an $s'$-vector basis $\ebf_J$ with degree $s'=\len{J}$, then
\begin{equation} \label{eq:ext-prod-def}
	\ebf_I\wedge\ebf_J = \sigma(I,J)\ebf_{I+J},
\end{equation}
indicating $\sigma(I,J)$ as the sign of the permutation given by sorting the elements of this concatenated list of $\len{I}+\len{J}$ indices, and $I+J$ define the resulting sorted list.

Apart from exterior product, we formalize the concept of interior product, too.
Let us consider two multivectors basis $\ebf_I$ and $\ebf_J$, such that $I$ is a subset of $J$, namely $I\subset J$. Then the left and right interior products, denoted respectively as $\lintprod$ and $\rintprod$, are defined
\begin{equation}	\label{eq:left-int-prod}
	\ebf_I \lintprod \ebf_J = \Delta_{II}\sigma(J\setminus I,I)\ebf_{J\setminus I},
\end{equation}
and
\begin{equation}\label{eq:right-int-prod}
	\ebf_J \rintprod \ebf_I = \Delta_{II}\sigma(I,J\setminus I)\ebf_{J\setminus I}, 
\end{equation}
where the multivector $\ebf_{J\setminus I}$ has grade $\len{J}-\len{I}$ and it contains the elements of $J$ not in common with $I$.
The term $\Delta_{II}$ comes out from the generalized metric, defined as $\Delta_{IJ}=\Delta_{i_1 j_1}\Delta_{i_2 j_2}\dots\Delta_{i_s j_s} $, where $\Delta_{ij}$ vanishes if $i\ne j$, $\Delta_{ii}=-1$ if $i=0,1,\dots , k-1 $, thus time-like, and $\Delta_{ii}=+1$ for space-like indices, indeed $i=k, \dots, k+n-1$.

It might be interesting to notice that the difference between left and right interior products is only given by a sign and it depends on the grade of the two multivectors~\cite{colombaro2019introductionSpaceTimeExteriorCalculus}, expressed as
\begin{equation}
\ebf_I \lintprod \ebf_J = \ebf_J \rintprod \ebf_I (-1)^{\len{I}(\len{J} - \len{I})},
\end{equation}
while in case $\len{I}= \len{J} = s$ we recover the expression for the generalized dot product in exterior calculus, which reads 
\begin{equation}
\ebf_I \lintprod \ebf_J = \ebf_I \rintprod \ebf_J = \ebf_I \cdot \ebf_J = \Delta_{IJ} , \qquad \len{I}= \len{J}.
\end{equation}

\subsection{Time-like formulation of quaternions in exterior algebra}

Let us setup a three-dimensional spacetime with time-like coordinates. As shown in~\cite[Sect.~3]{colombaro2023quaternions}, we refer to the coordinates as $\ebf_\iq$, $\ebf_\jq$ and $\ebf_\kq$, such that $\Delta_{\iq\iq}=\Delta_{\jq\jq}=\Delta_{\kq\kq}=-1$, and the so-called Grassmann or Hodge complement~\cite{Frankel2012geomphys} satisfies the following relations for vectors
\begin{gather}
\ebf_\iq^\hodge = - \sigma(\iq, \jq\kq) \ebf_{\jq\kq}
\\
\ebf_\iq^\hodgeinv = \sigma(\jq\kq,\iq) \ebf_{\jq\kq}	,\label{eq:vec-invhodge}
\end{gather}
and for bivectors
\begin{gather}
\ebf_{\iq\jq}^\hodge = \sigma(\iq\jq,\kq)\ebf_\kq \label{eq:bivec-hodge}
\\
\ebf_{\iq\jq}^\hodgeinv = - \sigma(\kq,\iq\jq)\ebf_\kq . \label{eq:bivec->vec}
\end{gather}

The time-like basis $\{\ebf_\iq, \ebf_\jq, \ebf_\kq\}$ with the addition of a scalar term builds objects as introduced in~\eqref{eq:introeq} and we name the set of these kinds of elements as $\Qset$. Indeed, elements $\ptf, \qtf \in \Qset$ satisfy some properties which are listed hereinafter. 
First, the generalized product between exterior-algebraic quaternionic elements, or simply $\Qset$-product, is defined as
\begin{equation}\label{eq:Qextprod}
    \ptf \wedgequat \qtf = \ptf \cdot \qtf + \ptf \wedgevec \qtf ,
\end{equation}
corresponding to the so called Hamilton product~\cite{colombaro2023quaternions}. 
A particular mention is needed for the vector wedge operator $\wedgevec$, which corresponds with the standard vector product for three dimensional vectors, and it comes from the definition of bivectors in the $(1,3)$-spacetime, written 
\begin{equation}\label{eq:bivec-vec}
\ebf_{uv} = \ebf_u \wedgevec \ebf_v = \sigma(uv,w)\ebf_w \,,
\end{equation}
where $u,\,v,\,w={\iq},\,{\jq},\,{\kq}$ and for $\sigma(uv,w)$ being the sign of the permutation to order the triplet $(u\,v\,w)$~\cite[Property~3.1]{colombaro2023quaternions}.
Defined the vectors $\pbf=(p_1, p_2, p_3)$ and $\qbf=(q_1, q_2, q_3)$ in the time-like basis triplet $\{\ebf_\iq, \ebf_\jq, \ebf_\kq\}$, we could also express 
$\ptf$ and $\qtf$ as $\ptf = p_0 + \pbf$ and $\qtf = p_0 + \qbf$, so that~\eqref{eq:Qextprod} might be written by means of the notation
\begin{equation}
    \ptf \wedgequat \qtf = p_0 q_0 + \pbf \cdot \qbf + q_0 \pbf + p_0 \qbf + \pbf \wedgevec \qbf ,
\end{equation}
or writing directly the vector wedge product $\wedgevec$ as the standard cross product between vectors
\begin{equation}
\ptf \wedgequat \qtf = p_0 q_0 - p_1 q_1 - p_2 q_2 - p_3 q_3 + q_0 \pbf + p_0 \qbf + \pbf \times \qbf .
\end{equation}

Secondly, there exist the conjugate of an object $\qtf\in \Qset$, represented by $\qtf^*$, where the vector elements present opposite sign, defined
\begin{equation}\label{eq:conjugatequat}
    \qtf^* = q_0 - q_1\ebf_\iq - q_2\ebf_\jq - q_3\ebf_\kq ,
\end{equation}
and, as a consequence, 
\begin{equation}
    (\ptf \wedgequat \qtf)^* = \qtf^* \wedgequat \ptf^* .
\end{equation}
Subsequently, the squared norm of $\qtf$ is defined
\begin{equation}\label{eq:Qnorm}
\vert \qtf \vert^2
= \qtf\wedgequat\qtf^*  = q_0^2 + q_1^2 + q_2^2 + q_3^2 ,
\end{equation}
in accordance with standard definition, and it follows that
\begin{equation}\label{eq:norm-gen-prod}
\vert \ptf \wedgequat \qtf \vert^2 = \vert \ptf \vert^2 \vert \qtf \vert^2 .
\end{equation}
Furthermore, utilizing the conjugate and the squared norm, it is possible to define the inverse of a quaternion $\qtf$ as
\begin{equation}\label{eq:def-inv-quat}
\qtf^{-1}= \frac{\qtf^{*}}{\vert \qtf \vert ^2 },
\end{equation}
and in this manner, having the identity operator $\mathbf{1}$, we have that
\begin{equation}\label{eq:inv-def-id}
\qtf^{-1} \wedgequat \qtf = \mathbf{1} =\qtf \wedgequat \qtf^{-1}.
\end{equation}

This framework is the general system considered for defining electromagnetism by means of exterior-algebraic quaternions.
Such a time-like representation of the quaternionic algebra in exterior calculus finds a few similarities with the representation described in~\cite{inoguchi1998timelikesurfaces}, where a geometrical interpretation is also provided.

\section{Exterior-algebraic quaternions and Maxwell equations}
\label{sec:ME}

In order to define quaternionic electromagnetic theory in exterior algebra, we have to define quaterions with imaginary scalar part and time-like vectorial part.
To this end, let define first the quaternion associated to the derivatives
\begin{equation}\label{eq:qdev}
    \Dq=\frac{\imath}{c} \partial_0 - \nabla    ,
\end{equation}
where $\nabla = \partial_1 \ebf_\iq +\partial_2 \ebf_\jq +\partial_3 \ebf_\kq$ contains the derivatives with respect to the three coordinates $(x_1, x_2, x_3)$ and $\partial_0$ represents the derivative with respect to a parameter $x_0$.
Let then consider the conjugate of the quaternionic spacetime derivative~\eqref{eq:qdev} as
\begin{equation} \label{eq:D*}
    \Dq^{*} = \frac{\imath}{c} \partial_0 + \nabla  ,
\end{equation}
recognizing that the conjugate is executed with respect to the vectorial part of the quaternionic basis and not with respect to the imaginary unit $\imath$.
It is then straightforward to prove that
\begin{equation} \label{eq:eaqbox}
    \Dq^{*} \wedgequat \Dq = \Dq \wedgequat\Dq^{*} = -\frac{1}{c^2}\partial^2_0 + \partial^2_1 +\partial^2_2 +\partial^2_3 = \square ,
\end{equation}
and we can point out the equivalence between the result obtained in~\eqref{eq:eaqbox} and the D'Alembert operator $\square$.
For completeness, in the context of exterior calculus,~\eqref{eq:eaqbox} corresponds to $\deltabf\cdot \deltabf$ for the spacetime coordinates $k=1$ and $n=3$~\cite[Sec.~2.3]{colombaro2020generalized}, so that in the equivalence between exterior-algebraic quaternionic electromagnetism and the classical formulation of Maxwell equation, $x_0$ corresponds to the time variable.

Subsequently, let define a quaternion $\Aq = \Aq(x_0, \xbf)$, depending on $x_0$ and the coordinates $\xbf=(x_1, x_2, x_3)$ associated to the scalar and vector potentials
\begin{equation}\label{eq:quatpot}
    \Aq = \frac{\imath}{c}\phi -\abf ,
\end{equation}
where the vectorial part $\abf(x_0, \xbf)$ of this latter quaternion is defined $\abf = A_1 \ebf_\iq + A_2 \ebf_\jq + A_3 \ebf_\kq$ and recalls the vector potential, while the scalar function corresponding to the scalar potential is $\phi(x_0, \xbf)$.

We can compute the $\Qset$-exterior product~\eqref{eq:Qextprod} between the quaternion derivative in~\eqref{eq:qdev} and the quaternion potential~\eqref{eq:quatpot}, namely
\begin{equation}\label{eq:DwqA}
    \Dq \wedgequat \Aq =
    \left( -\frac{1}{c^2} \partial_0\phi  + \nabla \cdot \abf\right) + \nabla \times \abf - \frac{\imath}{c}\left( \partial_0\abf  + \nabla \phi \right).
\end{equation}
From~\eqref{eq:DwqA}, we can notice that the first (scalar) summand on the right-hand side remind to the relation known as Lorenz gauge~\cite{jackson1999classicalElectrodynamics}, so that it would be equivalent to a gauge condition and it might be set to zero.
The second summand instead coincides with the magnetic flux density $\Bb = \nabla \times \abf$, given by the result of the curl of the equivalent vector potential, while the last term contains the expression for the electric field in the Lorenz gauge, namely $\Eb =-\partial_0\abf  - \nabla \phi $.

As a consequence, assuming the validity of the Lorenz gauge in order to have a vanishing scalar term, we can define the pure quaternion from~\eqref{eq:DwqA} as
\begin{equation}
    \Fq = \Bb +   \frac{\boldsymbol{\imath}}{c}\Eb   ,
\end{equation}
containing the two electric and magnetic fields, recovering an expression compatible with the results obtained in~\cite{girard1984quatgroup} and in regards of the Lorentz group.
Having then the conjugate $\Fq^* = -\Bb -   \frac{\boldsymbol{\imath}}{c}\Eb$, since $\Fq$ is a pure quaternion so it does not present any scalar term, we are able to define the norm of the electromagnetic quaternion
\begin{equation}
    \vert \Fq \vert^2 = \Fq \wedgequat \Fq^* = \vert \Bb \vert^2 - \frac{\vert \Eb \vert^2}{c^2} - \frac{2\imath}{c} \Eb \cdot \Bb  .
\end{equation}
In analogy with the electromagnetic multivector, given by the exterior product of the multivector potential~\cite{colombaro2020generalized}, we find that the electromagnetic quaternion is given by the $\Qset$-product between quaternion derivative and quaternion potential. 

For completeness, we define also the quaternion containing the so called external sources, namely
\begin{equation}
    \Jq = \imath \, c\rho + \jbf    ,
\end{equation}
where $\rho$ stands for the density of charge and $\jbf$ is the vector density of current, expressed in the time-like basis of our mathematical framework.
Thus, we compute
\begin{equation}\label{eq:nonhmax}
    \Dq^* \wedgequat \Dq \wedgequat \Aq =
    \Dq^* \wedgequat \Fq = \mu_0 \Jq    ,
\end{equation}
and we state the following conclusions. Applying the conjugate quaternionic spacetime derivative~\eqref{eq:D*} to $\Fq$, we are then able to isolate the real and imaginary parts, respectively written
\begin{gather}
\nabla \cdot \Bb - \frac{1}{c^2}\partial_0\Eb + \nabla \times \Bb = \mu_0 \jb, \label{eq:maxquat-step1re}
\\
\frac{1}{c} \nabla \cdot \Eb + \frac{1}{c}\partial_0 \Bb + \frac{1}{c}\nabla \cdot \Eb = \mu_0 c\rho, \label{eq:maxquat-step1im}
\end{gather}
and separating the scalar and vector parts in these latter~\eqref {eq:maxquat-step1re} and~\eqref {eq:maxquat-step1im}, we recover the classical expression of Maxwell equations as in~\eqref{eq:maxEq_intro_1},~\eqref{eq:maxEq_intro_2},~\eqref{eq:maxEq_intro_3} and~\eqref{eq:maxEq_intro_4}.

An interesting but still open problem is then related to the application of the operator $\Dq \wedgequat$ to~\eqref{eq:nonhmax}, i. e. $\Dq \wedgequat \Jq$, 
whose scalar part leads to the known expression of the continuity equation for the conservation of the charge, and whose vectorial imaginary and real parts recover the extra terms obtained in the evaluation of the divergence of the interaction stress-energy-momentum tensor~\cite[Sec.~4.3, Eq.~(194)]{martinez2021tensor}, that is still open for physical interpretation.

\section{Conclusions} \label{sec:last}

To finally recap the main results, we have been able to reacquire the homogeneous and non-homogeneous Maxwell equations of classical electromagnetism, expressed with the notation of generalized exterior calculus~\cite{colombaro2020generalized}, considering a spacetime $(1+3)$-dimensional, by means of time-like exterior-algebraic quaternions with complex scalar part.
For a given pure quaternion electromagnetic field $\Fq$ , which is related to the potential $\Aq$ as in~\eqref{eq:quatpot} by means of the relation
\begin{equation}
\Fq =   \Dq \wedgequat \Aq	,
\end{equation}
then both the inhomogeneous and homogeneous Maxwell equations are included in the single equation
\begin{gather}
\Dq^* \wedgequat \Fq = \mu_0 \Jq	\label{eq:quatMaxinh}
\end{gather}


Concerning electromagnetism, it is remarkable to highlight the similarities with the work of Salingaros. For example, the mathematical framework taken into account~\cite{salingaros19831} presents considerable peculiarities, as the creation of an equivalent electromagnetic theory~\cite{salingaros1981electromagnetism}.


In the framework of exterior calculus, Maxwell equations have been computed by simple analogy by simply evaluating interior and exterior derivatives of an $r$-vector field~\cite{colombaro2020generalized} and also by means of variational methods~\cite{colombaro2021EulerLagrange}, obtaining the corresponding result.
In the current work, the formula obtained in~\eqref{eq:quatMaxinh} is then equivalent with the corresponding expressions in exterior algebra~\eqref{eq:eamax-inh} and~\eqref{eq:eamax-h}, when the dimensions of the spacetime are fixed to $k=1$ and $n=3$. So, there exist a relation of equivalence between the exterior-algebraic multivectorial electromagnetism and the quaternionic formulation made starting from complex-scalar-term time-like quaternions.

\section*{Acknowledgments}


The author acknowledge the anonymous reviewer for the constructive comments and suggestions which have
helped to improve the manuscript.
The work of the author has been carried out in the framework of the activities of the Italian National Group of Mathematical Physics [Gruppo Nazionale per la Fisica Matematica (GNFM), Istituto Nazionale di Alta Matematica (INdAM)].

\bibliographystyle{IEEEtran}
\bibliography{physics-rdm}

\begin{thebibliography}{10}
\providecommand{\url}[1]{#1}
\csname url@samestyle\endcsname
\providecommand{\newblock}{\relax}
\providecommand{\bibinfo}[2]{#2}
\providecommand{\BIBentrySTDinterwordspacing}{\spaceskip=0pt\relax}
\providecommand{\BIBentryALTinterwordstretchfactor}{4}
\providecommand{\BIBentryALTinterwordspacing}{\spaceskip=\fontdimen2\font plus
\BIBentryALTinterwordstretchfactor\fontdimen3\font minus
  \fontdimen4\font\relax}
\providecommand{\BIBforeignlanguage}[2]{{%
\expandafter\ifx\csname l@#1\endcsname\relax
\typeout{** WARNING: IEEEtran.bst: No hyphenation pattern has been}%
\typeout{** loaded for the language `#1'. Using the pattern for}%
\typeout{** the default language instead.}%
\else
\language=\csname l@#1\endcsname
\fi
#2}}
\providecommand{\BIBdecl}{\relax}
\BIBdecl

\bibitem{PerezGracia2020}
A.~Perez~Gracia, \emph{Quaternions and Clifford Algebras}.\hskip 1em plus 0.5em
  minus 0.4em\relax Berlin, Heidelberg: Springer Berlin Heidelberg, 2020, pp.
  1--12.

\bibitem{Wells1980}
R.~O. Wells, \emph{Differential Analysis on Complex Manifolds}, ser. Graduate
  Texts in Math. 65, Springer-Verlag, Ed., 1980.

\bibitem{salamon1986diffgeomquatman}
S.~Salamon, ``Differential geometry of quaternionic manifolds,'' \emph{Annales
  scientifiques de l'\'{E}cole {N}ormale {S}up\'{e}rieure}, vol.~19, no.~1, pp.
  31--55, 1986.

\bibitem{hamilton1853LectureNotes}
W.~R. Hamilton, \emph{Lectures on quaternions}.\hskip 1em plus 0.5em minus
  0.4em\relax Dublin: Hodges and Smith, 1853.

\bibitem{girard1984quatgroup}
P.~R. Girard, ``The quaternionic group and modern physics,'' \emph{Eur. J.
  Phys}, no.~5, p.~25, 1984.

\bibitem{hamilton1844onquaternions}
W.~R. Hamilton, ``{LXXVIII}. {O}n quaternions; or on a new system of
  imaginaries in algebra,'' \emph{The London, Edinburgh, and Dublin
  Philosophical Magazine and Journal of Science}, vol.~25, no. 169, pp.
  489--495, 1844.

\bibitem{Ablamowicz2004Clifford}
R.~Ablamowicz, W.~E. Baylis, T.~Branson, P.~Lounesto, I.~Porteous, J.~Ryan,
  J.~M. Selig, and G.~Sobczyk, \emph{Lectures on Clifford (geometric) algebras
  and applications}, 2004th~ed., R.~Ablamowicz and G.~Sobczyk, Eds.\hskip 1em
  plus 0.5em minus 0.4em\relax Boston, MA: Birkhauser Boston, 2003.

\bibitem{Fraleigh2003algebra}
J.~B. Fraleigh, \emph{A First Course in Abstract Algebra}, 7th~ed.\hskip 1em
  plus 0.5em minus 0.4em\relax Reading, Massachusetts: Addison Wesley, 2003.

\bibitem{giardino2024diffgeom}
S.~Giardino, ``Differential geometry using quaternions,'' 2024.

\bibitem{Kuipers1999quaternions}
J.~B. Kuipers, \emph{Quaternions and Rotation Sequences}.\hskip 1em plus 0.5em
  minus 0.4em\relax Princeton: Princeton University Press, 1999.

\bibitem{Hearn1997ComputerGraphics}
D.~Hearn and M.~P. Baker, \emph{Computer Graphics}, 2nd~ed.\hskip 1em plus
  0.5em minus 0.4em\relax US: Prentice Hall, 1997.

\bibitem{colombaro2023quaternions}
I.~Colombaro, ``Time-like definition of quaternions in exterior algebra,''
  \emph{Ric. di Mat.}, vol.~73, pp. 2865--2876, 2024.

\bibitem{colombaro2019introductionSpaceTimeExteriorCalculus}
I.~Colombaro, J.~Font-Segura, and A.~Martinez, ``An introduction to space--time
  exterior calculus,'' \emph{Mathematics}, vol.~7, pp. 564--583, 2019.

\bibitem{dressel2015stalgebra}
J.~Dressel, K.~Y. Bliokh, and F.~Nori, ``Spacetime algebra as a powerful tool
  for electromagnetism,'' \emph{Physics Reports}, vol. 589, pp. 1--71, 2015,
  spacetime algebra as a powerful tool for electromagnetism.

\bibitem{colombaro2020generalized}
I.~Colombaro, J.~Font-Segura, and A.~Martinez, ``Generalized {M}axwell
  equations for exterior-algebra multivectors in $(k,n)$ space-time
  dimensions,'' \emph{Eur. Phys. J. Plus}, vol. 135, no. 305, 2020.

\bibitem{jackson1999classicalElectrodynamics}
J.~D. Jackson, \emph{Classical Electrodynamics}, 3rd~ed.\hskip 1em plus 0.5em
  minus 0.4em\relax New York: John Wiley \& Sons, 1999.

\bibitem{PDG2020}
P.~A. Zyla~\textit{et al.} (Particle Data~Group), \emph{Prog. Theor. Exp. Phys.
  2020, 083C01}, 2020.

\bibitem{Frankel2012geomphys}
T.~Frankel, \emph{The Geometry of Physics}, 3rd~ed.\hskip 1em plus 0.5em minus
  0.4em\relax Cambridge: Cambridge University Press, 2012.

\bibitem{inoguchi1998timelikesurfaces}
J.~Inoguchi, ``Timelike surfaces of constant mean curvature in {M}inkowski
  3-space,'' \emph{Tokyo J. Math.}, vol.~21, no.~1, pp. 141--152, 1998.

\bibitem{martinez2021tensor}
A.~Martinez, J.~Font-Segura, and I.~Colombaro, ``An exterior-algebraic
  derivation of the symmetric stress-energy-momentum tensor in flat
  space-time,'' \emph{Eur. Phys. J. Plus}, vol. 212, no. 136, 2021.

\bibitem{salingaros19831}
N.~Salingaros and M.~Dresden, ``Physical algebras in four dimensions. {I}. the
  {C}lifford algebra in {M}inkowski spacetime,'' \emph{Adv. Appl. Math.},
  vol.~4, no.~1, pp. 1--30, 1983.

\bibitem{salingaros1981electromagnetism}
N.~Salingaros, ``Electromagnetism and the holomorphic properties of
  spacetime,'' \emph{J. Math. Phys.}, vol.~22, pp. 1919--1925, 1981.

\bibitem{colombaro2021EulerLagrange}
I.~Colombaro, J.~Font-Segura, and A.~Martinez, ``An exterior algebraic
  derivation of the {E}uler--{L}agrange equations from the principle of
  stationary action,'' \emph{Mathematics}, vol.~9, no.~18, p. 2178, 2021.

\end{thebibliography}

\end{document}